\documentclass[aps,pre,twocolumn,showpacs,superscriptaddress]{revtex4}
\usepackage[latin1]{inputenc}
\usepackage[english]{babel}
\usepackage{bm,graphicx,graphics,amsmath,amssymb,epsfig,color}

\usepackage[colorlinks,citecolor=blue,linkcolor=cyan]{hyperref}
\usepackage[usenames,dvipsnames,svgnames,table]{xcolor}

\def \be {\begin{equation}}
\def \ee {\end{equation}}
\def \beA {\begin{eqnarray}}
\def \eeA {\end{eqnarray}}

\def \der {\partial}

\def \average#1{\left\langle #1 \right\rangle}
\def \roundb#1{\left( #1 \right)}

\begin{document}
\title{Gauge invariance and geometric phase in nonequilibrium thermodynamics}

\author{Simone Borlenghi} 
\affiliation{Department of Physics and Astronomy, Uppsala University, Box 516, SE-75120 Uppsala, Sweden.}

\begin{abstract}
We show the link between $U_1$ lattice gauge theories and the off-equilibrium thermodynamics of a large class of nonlinear oscillators networks. 
The coupling between the oscillators plays the role of a gauge field, or connection, on the network. The thermodynamical forces that drive energy flows are expressed in terms of the curvature
of the connection, analogous to a geometric phase. The model, which holds both close and far from equilibrium, predicts the existence of persistent energy and particle currents circulating in close loops through the network.
The predictions are confirmed by numerical simulations. Possible extension of the theory and experimental applications to nanoscale devices are briefly discussed.
\end{abstract}

\maketitle

%%%%%%%%%%%%
\section{introduction}%
%%%%%%%%%%%%

Gauge theories have played a prominent role in the development of modern Physics, spanning from the description of elementary particles to general relativity and condensed matter theory \cite{wilson74,kogut79,slavnov80,kogut83,kleinert89,hehl95}. 
However, the role of gauge invariance in the formulation of non-equilibrium thermodynamics is still poorly understood.

The standard approach to non equilibrium transport processes is formulated in terms of generalised forces and coupled currents \cite{onsager31a,onsager31b,kubo57a,kubo57b}. In the study of heat transport in nonlinear lattices \cite{lepri03,dhar08}, one typically 
considers a system connected to several thermal reservoirs. 

The equations of motion of the ensemble (system+reservoirs) are solved numerically as a function of different reservoir parameters, such as temperature and chemical potential. 

The relevant currents are calculated in the stationary state and related to the
thermodynamical forces by computing the Onsager matrix. The forces, which relate the dynamics to the thermodynamics of the system, are in general expressed as temperature and chemical
potential differences or gradients. The main limitation of this formulation consists in that it is valid close to thermal equilibrium.

A more general approach was developed by Schnakenberg \cite{schnakenberg76}. Here a Markovian process is described by a master equation and represented by a graph, where the nodes correspond to probabilities associated
to the states of the system and the links to transition probabilities (or amplitudes) between the states. Thermodynamical forces, that drive the probability currents in the system, are expressed in terms of close paths in the graph. From  
probability currents,  heat, energy and particle flows can be obtained. The formulation holds arbitrary far from equilibrium and applies both to classical and quantum master equations \cite{esposito06}.

Recently, M. Polettini has shown that the Schnakenberg model can be formulated as a gauge theory \cite{polettini12}, where thermodynamical forces  play the role of a gauge potential.

Other lines of research were focussed on the covariant formulation of the Fokker-Planck equation \cite{graham77,feng11} and on relating the Onsager reciprocity relations to global gauge symmetries \cite{gambar93}.

In the present Paper, we investigate the role of gauge symmetries in the transport properties of a network of nonlinear oscillators, modelled by a generalisation of the discrete nonlinear Schr\"odinger equation (DNLS) \cite{rasmussen00,kevrekidis01}.

The recent advances on the non equilibrium DNLS \cite{iubini12,iubini13,borlenghi14a,borlenghi14b,borlenghi15a} have shown that transport depends on the synchronisation and phase differences between the oscillators,
sharing strong similarities with the well known Josephson effect \cite{josephson62}. 

In the present work, we formalise those observations by showing that the coupling between the oscillators plays the role of a gauge field, or connection, on the network. The thermodynamical forces that drive energy flows are expressed in terms of the curvature
of the connection, analogous to a geometric phase. This ultimately stems from the fluctuation-dissipation theorem and from the $U_1$ gauge invariance of the DNLS.

\begin{figure}
\begin{center}
\includegraphics[width=0.99\columnwidth]{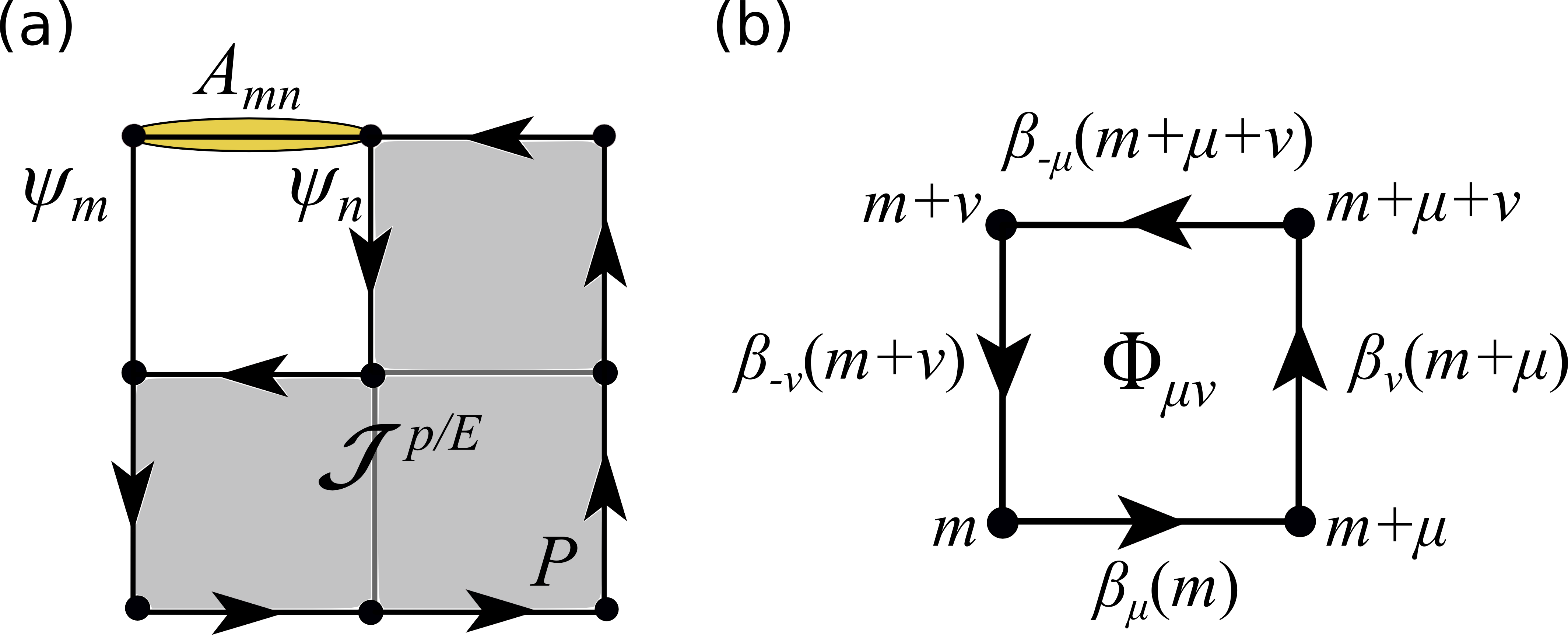}
\end{center}
\caption{(Color online) a) Cartoon of an oscillator network described in terms of nodes $\psi_m$ and links $A_{mn}=C_{mn}e^{i\beta_{mn}}$. The direction of the current is indicated by the black arrows. The phase $\beta_{mn}$ controls persistent particle and energy currents $\mathcal{J}^{p/E}$ 
along a close path $P$ on the network (shaded grey area).  b)  Plaquette of four oscillators, described in terms
of nodes and links coordinates. The link $(m,\mu)$ connects the node $m$ to its neighbour $n=m+\mu$ in the $\mu$ direction. The phase $\beta_{mn}$, which in this notation reads $\beta_{\mu}(m)$, plays the role of a $U_1$ gauge field, similar to a vector potential.
The total phase $\Phi_{\mu\nu}$ around the plaquette is the thermodynamical force that drives circulating currents, similar to a magnetic field (see text).} 
\label{fig:figure1}
\end{figure}

Our model predicts in particular the existence of persistent energy and particle currents circulating in close loops through the network, as shown in Fig.\ref{fig:figure1}a).

The present Paper is organised as follows. In Sec. II we review transport in the DNLS model. Sec. III contains the main result of the paper, where we show the link between the off-equilibrium thermodynamics of 
oscillator networks and lattice gauge theory. Here we identify thermodynamical forces with lattice gauge fields and describe circulating currents.  Sec. IV contains numerical simulations that further inspect and corroborate the model. 
The conclusions, together with a summary of the main results and possible developments, are reported in Sec. V. 

%%%%%%%%%%%%%%%%%%
\section{Oscillator network model}%
%%%%%%%%%%%%%%%%%%

%%%%%%%%%

The dynamics of several physical systems, such as Bose-Einstein condensates, photonics waveguides, lasers, mechanical oscillators, spin and electronic systems \cite{kevrekidis01,eilbeck03,rumpf03,slavin09,miller09,borlenghi14a,borlenghi15b}, can be described by the following universal oscillator model: 
\begin{align}
i\dot{\psi}_m = & \omega_m(p_m)\psi_m-i\Gamma_m(p_m)\psi_m+i\gamma_m\psi_m
\nonumber\\&
	+\sum_{n} A_{mn}\psi_n+ i\xi_m(t).
\label{eq:dnls}
\end{align}
Eq.(\ref{eq:dnls}) generalises the DNLS to a network of nonlinear oscillators with complex amplitudes $\psi_m=\sqrt{p_m(t)}{\rm{e}}^{i\phi_m(t)}$ and the geometry specified by the coupling $A_{mn}=C_{mn}{\rm{e}}^{i\beta_{mn}}$. 

The first term on the right hand side of Eq.(\ref{eq:dnls}) is the nonlinear frequency, while $\Gamma_m(p_m)=\alpha_m\omega_m(p_m)$ is the nonlinear damping, proportional to the parameter $\alpha_m$. 
Thermal baths are described in terms of the complex Gaussian random variable  $\xi_m(t)$
with zero average and variance $\average{\xi_m^*(t)\xi_n(t^\prime)}=D_m(p_m)T_m\delta_{mn}\delta(t-t^\prime)$ \cite{slavin09,iubini13,borlenghi14b}. The nonlinear diffusion constant $D_m(p_m)=\lambda\Gamma_m(p_m)/\omega_m(p_m)$ is prescribed by the fluctuation-dissipation theorem \cite{iubini13,slavin09}. $T_m$ is the bath temperature and $\lambda$ is a parameter that depends on the geometry of the oscillators \cite{slavin09}. The chemical potential $\gamma_m$ controls the relaxation time towards the reservoirs by compensating the damping.

Physical insight is gained by writing Eq.(\ref{eq:dnls}) in the phase-amplitude representation as \cite{slavin09,iubini13,borlenghi14a,borlenghi14b}. 
\begin{align}
\dot{p}_m = & 
	2[\gamma_m-\Gamma_m(p_m)]p_m+2\sqrt{p_m}{\rm{Re}}[\tilde{\xi}_m(t)]
\nonumber\\&
        +2D_m(p_m)T_m+\sum_n j^p_{mn}
\label{eq:amplitude}
\\
\dot{\phi}_m = &
	-\omega_m(p_m)+\frac{1}{\sqrt{p_m}}{\rm{Im}}[\tilde{\xi}_m(t)]
\nonumber\\&
       - \sum_{n}C_{mn}\sqrt{\frac{p_n}{p_m}}\cos(\Delta_{mn}+\beta_{mn}).
\label{eq:phase}
\end{align}
Eq.(\ref{eq:amplitude}) is the continuity equation that relates the time evolution of the oscillator power $p_m=|\psi_m|^2$ to the "particle" current
\be\label{eq:jp}
j_{mn}^p=2{\rm{Im}}[A_{mn}\psi_m^*\psi_n] 
\ee
between oscillators $(m,n)$. The ensemble-averaged current, which expresses the statistical correlation between the oscillators, can describe various transport processes, such as the particle flow in Bose-Einstein condensates and
the spin wave current in magnetic systems. 

Eq.(\ref{eq:phase}) describes the dynamics of the phases $\phi_m(t)$, whose evolution depends on the local nonlinear frequencies, on the relative powers and on the phase difference $\Delta_{mn}(t)=\phi_m(t)-\phi_n(t)$. 

In both Eqs.(\ref{eq:amplitude}) and (\ref{eq:phase}), temperature is described by means of the new random variable $\tilde{\xi_m}=e^{i\phi_m}\xi_m$, which has the same statistical properties
as $\xi_m$. Upon ensemble averaging, the stochastic terms vanish and the action of the bath appears through $2D_m(p_m)T_m$ in Eq.(\ref{eq:amplitude}). This source term ensures that the powers $p_m$ are never zero at
finite temperature. In non equilibrium steady states, $\dot{p}_m$ is zero in average and Eq.(\ref{eq:amplitude}) becomes 
\be
p_m=\frac{2D_m(p_m)T_m+\sum_n j^p_{mn}}{2(\Gamma_m(p_m)-\gamma_m)},
\ee
which shows that the power depends on the balance between sources, losses and currents. The values of the currents depend on the distribution of  temperature and chemical potential $(T_m,\gamma_m)$. 
When those are uniform, the system reaches thermal equilibrium, where currents vanish and the following equipartition relation holds: 

\be
p_m=\frac{D_m(p_m)T}{\Gamma_m(p_m)-\gamma}.
\ee
The model Eq.(\ref{eq:dnls}) contains another current. This can be seen by noting that the conservative part of Eq.(\ref{eq:dnls}) is given by the functional derivative  $\dot{\psi}_m=-i{\delta\mathcal{H}}/{\delta\psi_m^*}$ of the Hamiltonian \cite{slavin09,iubini13}

\be\label{eq:hamiltonian}
\mathcal{H} = {\sum_{m,n}}[\omega_m(p_m)p_m+A_{mn}\psi_m^*\psi_n +c.c.]. 
\ee
Computing the time evolution of $\mathcal{H}$  gives the energy current \cite{lepri03,iubini12,iubini13,borlenghi14a,borlenghi14b} 

\be\label{eq:je}
j_{mn}^E=2{\rm{Re}}[A_{mn}\dot{\psi}_m^*\psi_n].
\ee
Close to thermal equilibrium, the two coupled currents Eqs.(\ref{eq:jp}) and (\ref{eq:je}) are related to the thermodynamical forces (differences of temperatures and chemical potentials) through the Onsager matrix, according to the standard formulation of non equilibrium thermodynamics \cite{iubini12}. 
In the stationary state, the currents are conveniently written in the phase-amplitude representation as 
\beA
j_{mn}^p  &=& 2C_{mn}\sqrt{p_mp_n}\sin(\Delta_{mn}+\beta_{mn})\label{eq:jpphase},\\
j_{mn}^E  &=& 2C_{mn}\sqrt{p_mp_n}\omega_m(p_m)\sin(\Delta_{mn}+\beta_{mn})\label{eq:jephase}.
\eeA
From Eqs.(\ref{eq:jpphase}) and (\ref{eq:jephase}) one can see that the currents can flow whenever the oscillators are phase locked, so that $\Delta_{mn}$ reaches a constant value. The latter depends on the distribution of temperatures and chemical potentials trough Eqs.(\ref{eq:amplitude}) and (\ref{eq:phase}). If the oscillators are not phase-locked, $\Delta_{mn}$ fluctuates in time, so that the currents oscillate around zero and vanish in average \cite{borlenghi14a,borlenghi14b}.

From the fluctuation dissipation theorem follows that, at uniform $(T,\gamma)$ thermal equilibrium can be attained only if the coupling is \emph{dissipative}: $A_{mn}=\mathcal{A}_{mn}[1-i\Gamma_m(p_m)]$, where $\mathcal{A}_{mn}$ is a real matrix \cite{iubini13}. 

This amounts to fixing 
\be\label{eq:betaeq}
\beta_{mn}=\arcsin \roundb{\frac{\Gamma_m}{\sqrt{1+\Gamma_m^2}}}.
\ee
In this case, $\beta_{mn}\equiv\beta_{mm}$ is a \emph{local} quantity that contains only the $m$ index. 
Setting a \emph{nonlocal} $\beta_{mn}$, that depends on two oscillator indexes (or, in other terms, on the link between oscillators $m$ and $n$) drives the system out of equilibrium.

It has been recently shown by the present author that $\beta_{mn}$  acts as an additional thermodynamical force  that drives the current, in a way similar to the Josephson effect \cite{borlenghi15a} In particular, it permits to propagate persistent currents between oscillators at the same temperature, of from the colder to the hotter oscillator, operating the system as a heat pump. 

The relation between the locality/nonlocality of $\beta_{mm}$ and thermodynamical forces will be thoroughly discussed in the next section.

%%%%%%%%%%%%%%%%%%%%%%%
\section{connection with lattice gauge theory}%
%%%%%%%%%%%%%%%%%%%%%%%

\begin{figure}
\begin{center}
\includegraphics[width=0.8\columnwidth]{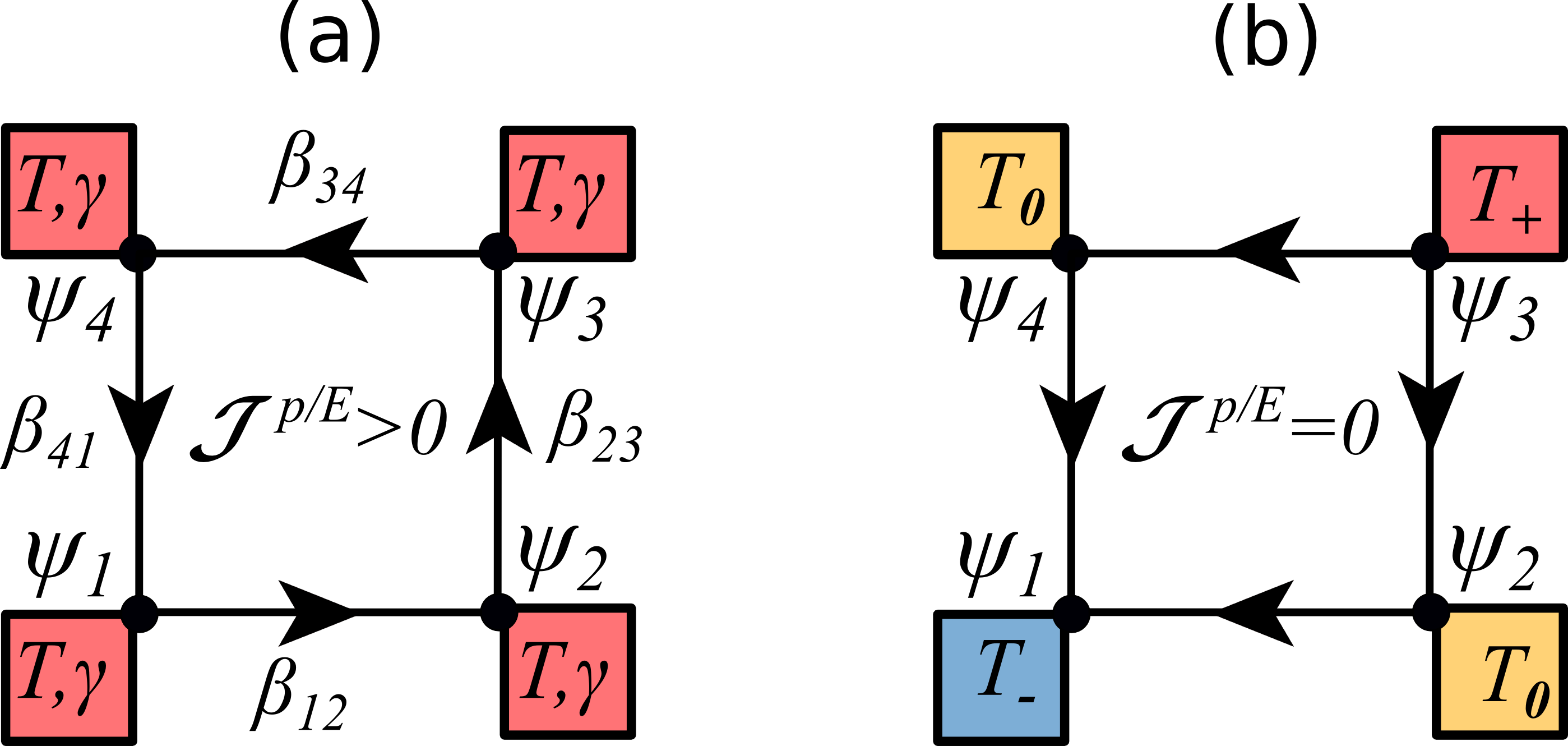}
\end{center}
\caption{(Color online) Cartoon showing four oscillators (represented by the dots) coupled to thermal baths (represented by the coloured squares at the corners). The direction of the currents is indicated by the black arrows. a) Persistent currents $\mathcal{J}^{p/E}$ circulating between oscillators at the same temperature and chemical potential $(T,\gamma)$, controlled by $\beta_{mn}$. Counterclockwise (resp. clockwise) currents are considered positive (resp. negative). b) Currents controlled solely by the differences between temperatures $T_+>T_0>T_-$. Those currents do not circulate, but flow from the hotter ($\psi_3$) to the colder ($\psi_1$) oscillator.} 
\label{fig:figure2}
\end{figure}

The possibility to control transport by tuning $\beta_{mn}$ suggests  that it should be possible to propagate persistent currents  $\mathcal{J}^{p/E}=\sum_P j_{mn}^{p/E}$ circulating along a close path $P$ in the oscillator network,
as shown in Figs.\ref{fig:figure1}a) and \ref{fig:figure2}a).

%%%%%%%%%

This unusual \emph{heat circuit}, where energy propagates in a close loop, might seems a paradox, since flows, described phenomenologically by thermodynamical forces, are due to the inhomogeneous spatial distribution of some physical quantities, such as concentrations and kinetic energy. 
The latter case is depicted in Fig.\ref{fig:figure2}b), where the currents generated by a temperature difference flow from the hotter ($T_+$) to the colder ($T_-$) oscillator, and there there is no circulating current.

From Eqs.(\ref{eq:jpphase}) and (\ref{eq:jephase}), it appears that persistent circulating currents require that the total phase $\Phi=\sum_{mn\in P}\beta_{mn}$ accumulated along a close path $P$ (or anholonomy angle) is non zero.
Here we shall see that $\beta_{mn}$ plays the role of a Gauge field and $\Phi$ is its curvature, that corresponds to the thermodynamical force that drives circulating currents.

Let us start from the stochastic, complex Lagrangian of the problem, which describes the ensemble (system+reservoirs).

\beA\label{eq:lagrangian}
\mathcal{L} &=& \frac{1}{2}\sum_{m,n}\{i(\dot{\psi}_m\psi_m^*-\dot{\psi}_m^*\psi_m)+[\omega_m(p_m)-i\Gamma_m(p_m)]p_m\nonumber\\
                   &-& (A_{mn}\psi^*_m\psi_n+A_{nm}\psi_m\psi_n^*)\nonumber\\
                   &+&i\sqrt{D_m(p_m)T_m}(\xi_m\psi_m^*-\xi_m^*\psi_m)\}.
\eeA
From the variation of the action $\delta\int\mathcal{L}dt=0$, one gets the Euler-Lagrange equations $\frac{d}{dt}\frac{\der\mathcal{L}}{\der{\mathcal {{\dot\psi}}_m^*}}-\frac{\der \mathcal{L}}{\der{\psi_m^*}}=0$,
which correspond to Eq.(\ref{eq:dnls}). The imaginary and stochastic parts of the Lagrangian account respectively for dissipation and thermal baths. The coupling matrix $A_{mn}$ in general is not Hermitian.

The Lagrangian Eq.(\ref{eq:lagrangian}) is unchanged by the global $U_1$ rotation $\psi_m\rightarrow e^{-i\alpha}\psi_m$. Thus, it is left invariant by the infinitesimal transformation $\delta\psi_m\approx-i\alpha\psi_m$ if and only if it satisfies the equation
$\delta \mathcal {L}=\sum_m\roundb{\frac{\der\mathcal{L}}{\der\dot{\psi}_m}\delta\dot{\psi}_m+\frac{\der\mathcal{L}}{\der{\psi}_m}\delta\psi_m+\frac{\der\mathcal{L}}{\der\dot{\psi}^*_m}\delta\dot{\psi}^*_m+\frac{\der\mathcal{L}}{\der\psi^*_m}\delta\psi^*_m}=0.$

A straightforward calculation shows that the latter is precisely the continuity equation Eq.(\ref{eq:amplitude}). In particular, $p_m$ corresponds to the Noether charge and $j^p_{mn}$ to the associated current.

A crucial point here is that the presence of the stochastic term $\xi_m$ does not break the $U_1$ symmetry \emph{in average}, since the statistical properties of the bath are unchanged by phase transformations. 

Equations and (\ref{eq:lagrangian}) is also left invariant by the more general, local gauge transformation  
\beA\label{eq:gauge}
\psi_m & \rightarrow & e^{-i\lambda_m}\psi_m,\nonumber\\
\psi_n^* & \rightarrow & e^{+i\lambda_n}\psi_n^*,\nonumber\\
\beta_{mn} & \rightarrow & \beta_{mn}-(\lambda_n-\lambda_m).
\eeA
Physically, this states the equivalence between the different reference frames associated to the oscillators. In other terms, the initial phases $\lambda_m$s of the $\psi_m$s can be arbitrarily chosen, but they have to 
be compensated a by proper rescaling of the coupling to leave the dynamics invariant. 

All the physical observables have to be gauge invariant. Together with the  local powers and currents $(p_m,j_{mn}^{p/E})$ the anholonomy angle 
$\Phi$ is invariant, since the sum $\sum_{mn\in P}(\lambda_m-\lambda_n)$ vanishes along a close path $P$.

The meaning of $\Phi$ becomes clear when we consider the simplest realisation of 2-dimensional lattice shown in Fig. \ref{fig:figure1}b), consisting of four coupled oscillators and called \emph{plaquette}. More complicate lattices can be constructed by joining several plaquettes.
By adopting the notation commonly used in lattice gauge theories \cite{wilson74,kogut79}, we describe the plaquette in terms of its nodes $m$, and links between the nodes along the coordinate directions $(\mu,\nu)$.

We label with the indexes $(m,\mu,\nu)$ the directed link between oscillator $m$ and its neighbour along the $\mu$ or $\nu$ direction. Using this notation, the coupling matrix reads $A_\mu=C_\mu e^{i\beta_\mu(m)}$, where  $\beta_{\mu}(m)$ corresponds to the phase $\beta_{mn}$ that connects oscillators $m$ and $n=m+\mu$. 
By introducing the finite difference gradient operator $\nabla_\mu\beta(m)=\beta(m+\mu)-\beta(m)$ and upon defining $\beta_{-\mu}(m+\mu)=-\beta_\mu (m)$ \cite{kogut79}, the anholonomy angle around the plaquette reads 

\be\label{eq:curl}
\Phi\equiv\Phi_{\mu \nu}(m)=\nabla_\mu\beta_\nu(m)-\nabla_\mu\beta_\mu(m),
\ee
which is the discretised version of the curl operator. Using the same notation, the gauge transformation Eq.(\ref{eq:gauge}) reads $\beta_\mu(m)\rightarrow\beta_\mu(m)-\nabla_\mu\lambda(m)$. 

Here $\beta_\mu(m)$ plays the role of a gauge field (connection) on the oscillator lattice, analogous to the vector potential of electromagnetism. Similarly, the total phase $\Phi_{\mu\nu}$ is the curvature of the connection, analogous to the Faraday field tensor. 
This is equivalent to a geometric phase, that depends only on the geometric properties of the path and not on the dynamics. 
On the other hand, the phase-differences $\Delta_{mn}$ between the oscillators in Eq.(\ref{eq:amplitude}) can be seen as dynamical phases, that depend on all the parameters $(T_m,\gamma_m,\beta_{\mu}(m))$, thus on the system dynamics. 
They vanish on close loops and therefore do not contribute to current circulation.

At thermal equilibrium, the phases are local quantities $\beta_m$ that obey Eq.(\ref{eq:betaeq}). This guarantees that the anholonomy angle $\Phi$ is zero, since Eq.(\ref{eq:curl}) reduces to the sum $\sum_{m,n}(\beta_m-\beta_n)$ which vanishes on close paths.
On the contrary, one can have a situation where there the system is out of equilibrium, but $\Phi=0$. In this case the local currents $j^{p/E}_{mn}$ between neighbouring oscillators do not vanish, but their sum around the path is zero,
so that there is no net circulating current.  

Gauge invariants can be in general expressed in terms of Wilson loops, i.e. path-oriented exponentials $W=P\exp\{\sum_{\mu\in P}\beta_\mu\}$, proportional to path ordered products of $A_\mu$.  
The latter can be also identified as the the gauge field instead of $\beta_\mu$. The Noether currents are gauge invariants given by the coupling of the gauge field  $A_{\mu}$ with the oscillator wavefunctions $(\psi_m,\psi_n)$,
that play the role of "matter fields" \cite{kogut79}.

%%%%%%%%%%%%%%%%%
\section{numerical simulations} %
%%%%%%%%%%%%%%%%%

To corroborate the model, we turn now to numerical simulations. 
We consider the off-equilibrium dynamics of a plaquette, made of the coupled oscillators $\psi_n$, $n=1,...,4$ shown in Fig.\ref{fig:figure2}a).

The oscillators have the same nonlinear frequencies and damping rates, respectively $\omega_n(p_n)=\omega^0\times(1+2p_n)$ and $\Gamma_n(p_n)=\alpha\omega_n(p_n)$ (in units where $k_B=1$).
Each oscillator is connected to an independent Langevin bath. The baths have temperatures and chemical potentials $(T_m,\gamma_m)$, and the same coupling strength $D=2\alpha$. 

Throughout all this section, for simplicity we adopt the notation of Sec.II and Fig. \ref{fig:figure2}a), with oscillator indexes $(m,n)$.
The coupling between oscillators $(m,n)$ reads then$A_{mn}=C(1-i\alpha){\rm{e}}^{i\beta_{mn}}$. The anholonomy angle, or gauge field curvature, reads $\Phi=\beta_{12}+\beta_{23}+\beta_{34}+\beta_{41}$.

The equations of motion were solved numerically using a fourth order Runge-Kutta algorithm with a time step ${\rm{d}}t=10^{-3}$. The other parameters adopted in all our simulations are $\omega^0=1$ and $C=0.1$. All 
the parameters are expressed in model units.
%%%%%%%%%
\begin{figure}
\begin{center}
\includegraphics[width=0.9\columnwidth]{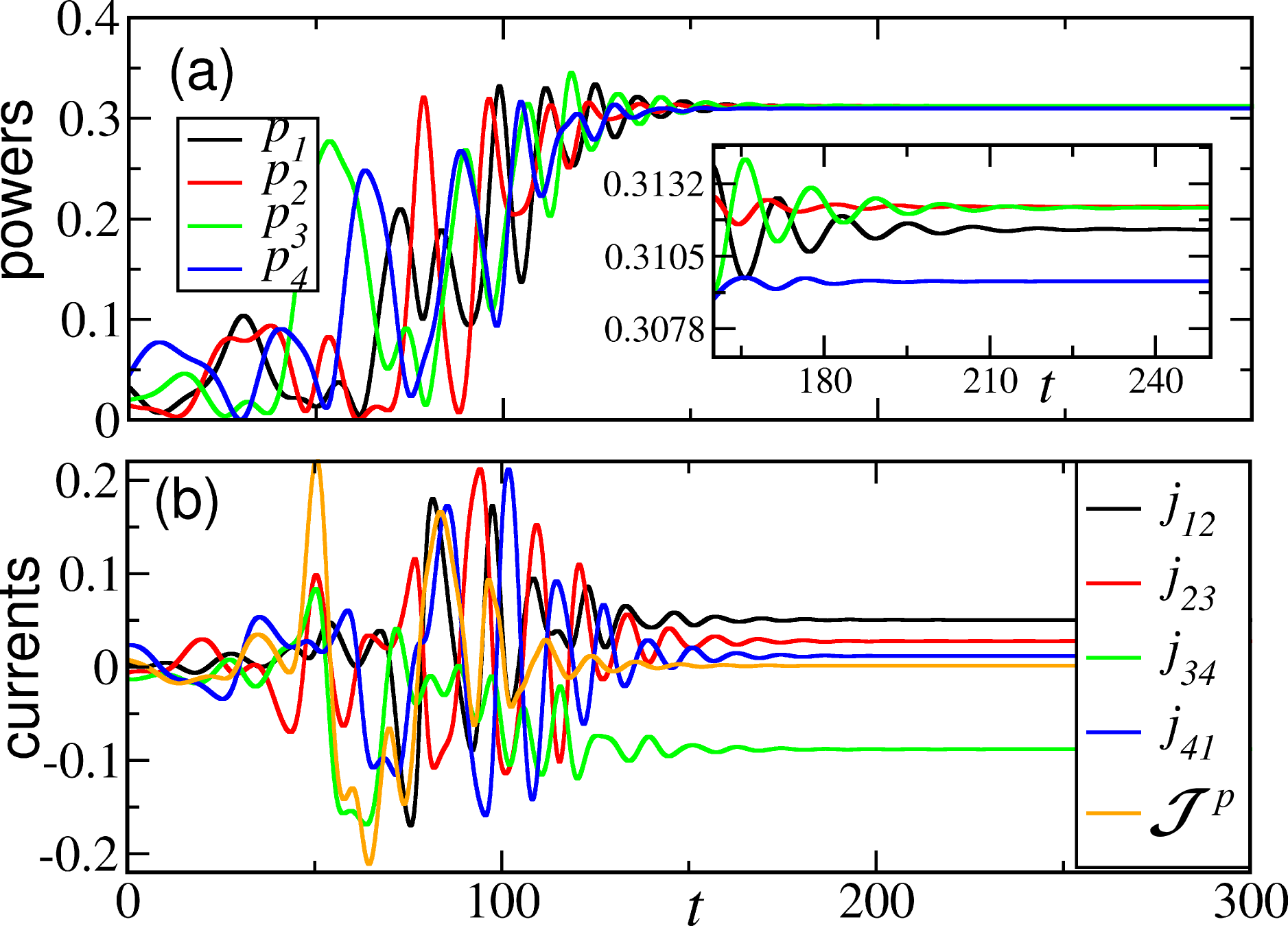}
\end{center}
\caption{(Color online). Time evolution of the system at zero temperature and phase $\beta_{mn}$. The off-equilibrium dynamics is given by a nonuniform distribution of chemical potential $\mu_m$, $m=1,...,4$. Panel (a) shows the dynamics of the powers $p_n$,
that reach constant values in the stationary state. The inset shows a detail of the dynamics in the stationary state, where one can see the difference between the powers. In panel (b) the time evolution of the particle currents $j_{mn}^p$ is displayed. The currents are nonzero, but their sum along the path $\mathcal{J}^p$ (displayed in orange tones) vanishes as expected} 
\label{fig:figure3}
\end{figure}

We consider first the case at zero temperature, and relatively high damping $\alpha=0.1$. For simplicity only the particle currents are reported, since the energy currents have the same profiles up to a scaling factor. 

Figs.\ref{fig:figure3} (a) and (b) show respectively the time evolution of the powers $p_m$ and of the particle currents $(j_{mn}^p, \mathcal{J}^p)$. Here $\beta_{mn}=0$ and the following values of chemical potential where chosen:
 $(\gamma_1=0.13,\gamma_2=0.15,\gamma_3=0.18,\gamma_4=0.11)$. After a transient regime the system reaches a non equilibrium stationary state where powers are different and local currents $j_{mn}^p$ are constant in time. 
 However, as expected from our model, the circulating current along the plaquette $\mathcal{J}^p$  (denoted in orange tones) is zero.

\begin{figure}
\begin{center}
\includegraphics[width=0.9\columnwidth]{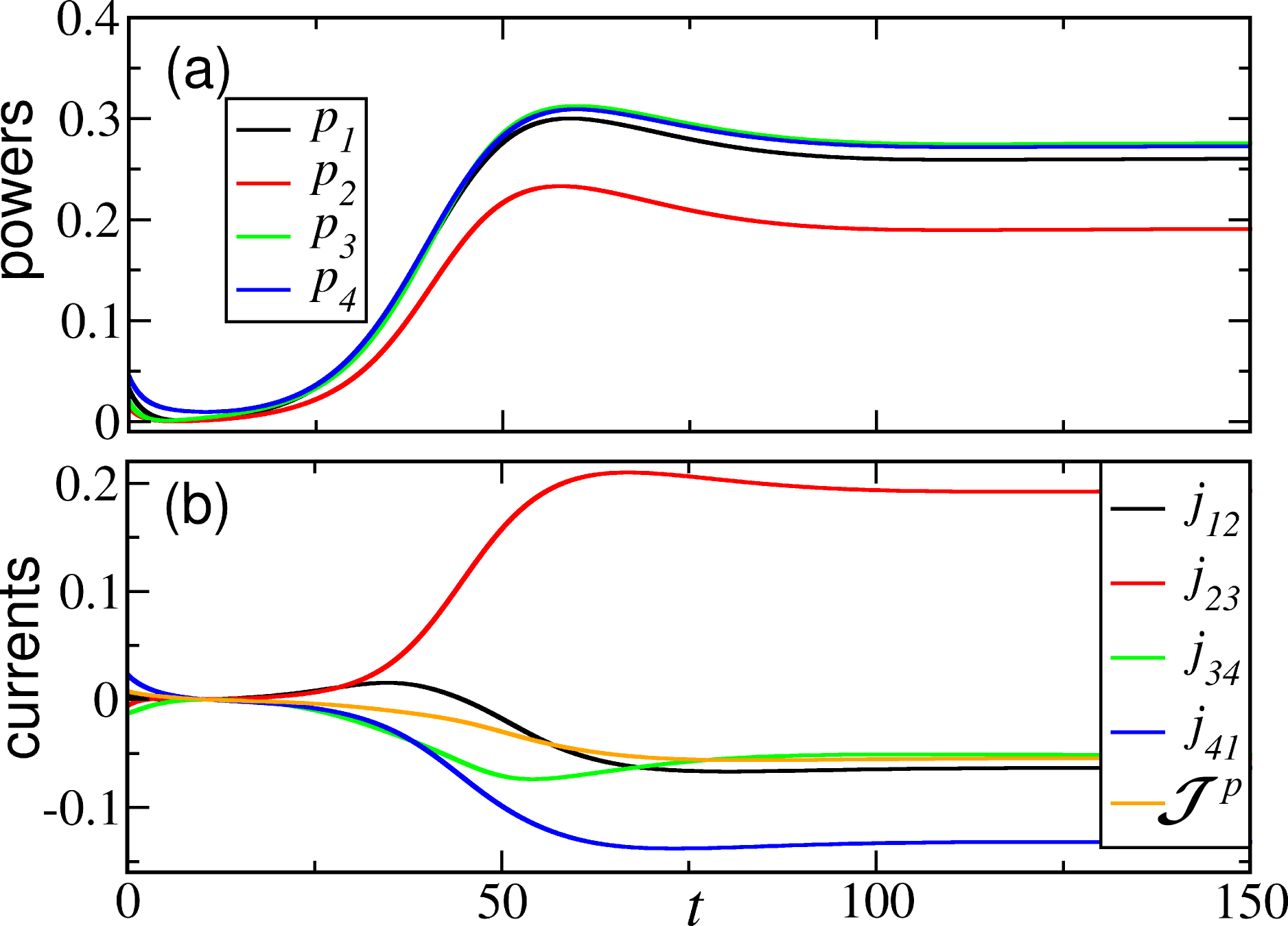}
\end{center}
\caption{(Color online). Time evolution of the system at zero temperature and chemical potential. The system is driven off-equilibrium solely by the phases $\beta_{mn}$. Panel (a) shows the dynamics of the powers $p_n$,
that reach constant values in the stationary state. Panel (b) shows the time evolution of the particle currents $j^p_{mn}$. Their sum along the path $\mathcal{J}^p$ (displayed in orange tones) does not vanish, since the anholonomy angle $\Phi$ is nonzero.} 
\label{fig:figure4}
\end{figure}

Figs.\ref{fig:figure4} (a) and (b) again display respectively the time evolution of powers and particle currents. In this case all chemical potentials are set to zero, while $\beta_{mn}$ has the following values:
$(\beta_{12}=0.25\pi,\beta_{23}=0.57\pi,\beta_{34}=0.38\pi,\beta_{41}=0.32\pi)$ These phases drive the system in a non equilibrium state where all the currents, including the sum around the plaquette $\mathcal{J}^p$ are nonzero.

%%%%%%%%%

Next, we consider the more general case where both temperature and chemical potentials are nonzero. 
In the following simulations the observables were time averaged in the stationary state over an interval of $3.5\times 10^6$ time steps and then ensemble-averaged over 100 samples with 
different realisations of the thermal field. 

We have taken a smaller damping parameter $\alpha=0.02$, and we have set $(\beta_{12}=0.5\pi, \beta_{23}=1.2\pi, \beta_{34}=0.8\pi)$. The curvature reads then $\Phi=0.49\pi+\beta_{41}$.
The dynamics was computed as a function of $\beta_{41}$, in the presence of uniform temperature and chemical potential $(T,\gamma)$.

%%%%%%%%%
\begin{figure}
\begin{center}
\includegraphics[width=0.9\columnwidth]{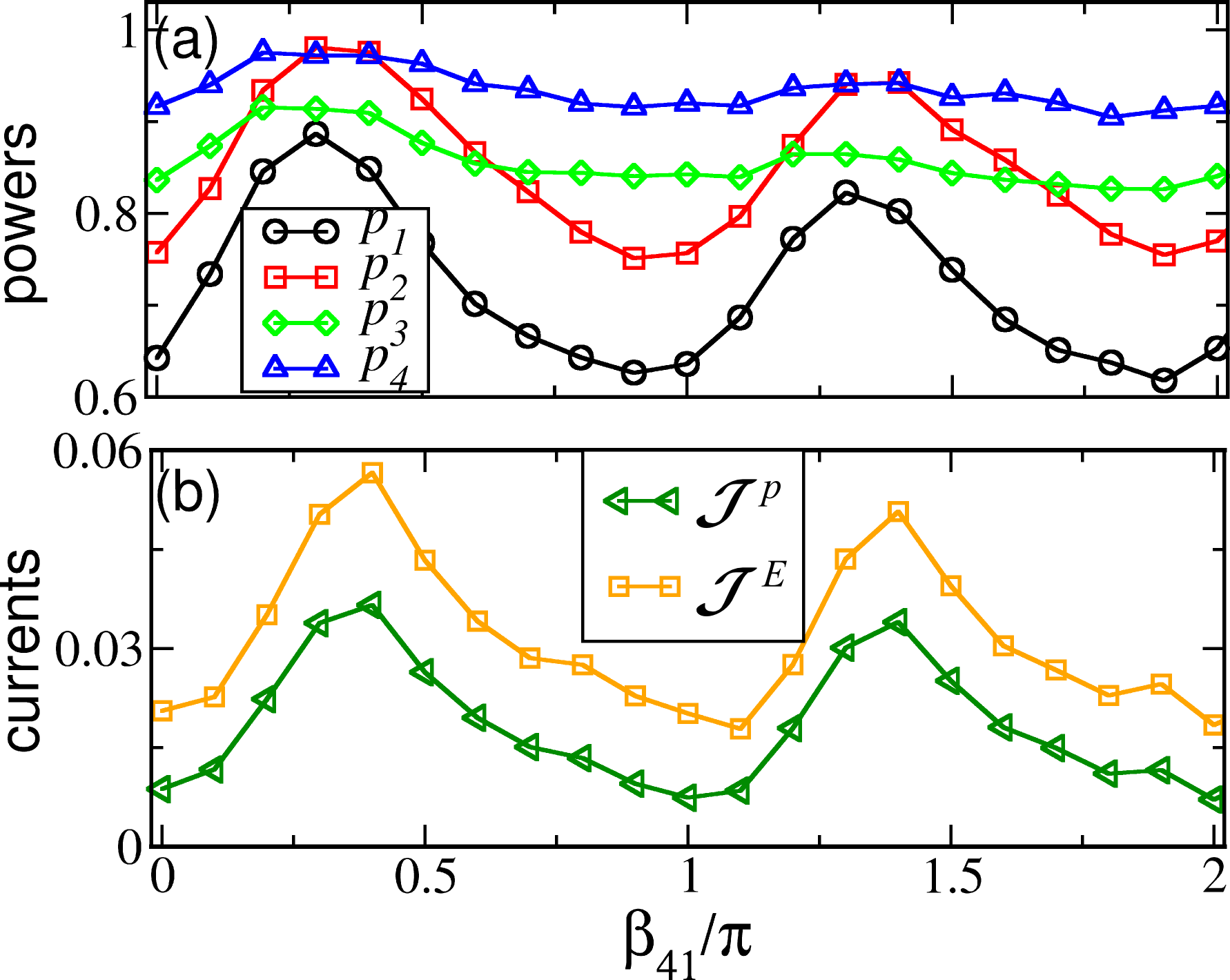}
\end{center}
\caption{(Color online) a) Oscillator powers and b) total currents vs $\beta_{41}$, computed using the bath parameters $T=0.3$ and $\gamma=0.021$ model units. The current $\mathcal{J}^p$ is magnified by a factor 3 for better visibility. 
The solid lines are guides to the eye.} 
\label{fig:figure5}
\end{figure}

Fig.\ref{fig:figure5} (a) and (b) shows respectively the powers and currents vs $\beta_{41}$. The computation where performed with bath parameters $T=0.3$ and $\gamma=0.021$.
One can see clearly that persistent currents circulate through the system. When the currents change, energy is redistributed through the system and also the local powers change. This testifies that the current circulation corresponds to a real transport 
process through the system. Apart from their different magnitudes, the currents $\mathcal{J}^{p/E}$ have similar profiles.

%%%%%%%%%
\begin{figure}
\begin{center}
\includegraphics[width=0.9\columnwidth]{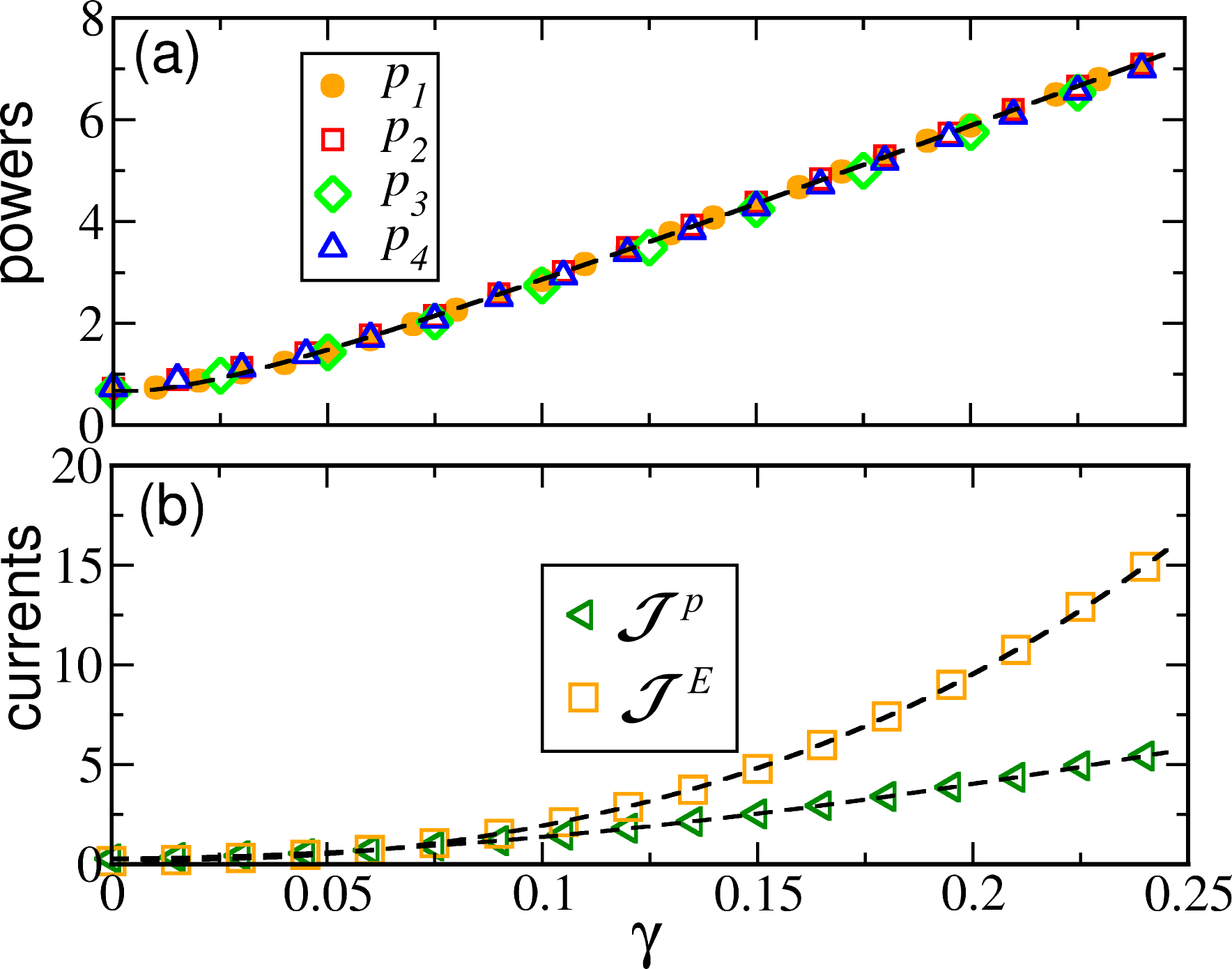}
\end{center}
\caption{(Color online) Oscillator powers (a) and currents (b) vs chemical potential $\gamma$, computed for $\beta_{41}=0.3\pi$ and $T=0.3$. Powers and currents grow with $\gamma$, as the energy of the system increases. 
The current $\mathcal{J}^p$ is magnified by a factor 6 for better visibility. The dashed black lines are fits with a Pad\'e function (see text). Since the powers have very close values, for better visibility in panel (a) only the fit for $p_1$ is displayed.} 
\label{fig:figure6}
\end{figure}

By increasing $(T,\gamma)$, the powers and energy of the system grow. One expect that this leads also to an increase in the currents.
This is indeed the case when $\gamma$ increases and $T$ is kept fixed, as one can see in Fig.\ref{fig:figure4} (a) and (b). Here powers and currents vs $\gamma$ are displayed, with parameters $T=0.03$ and $\beta_{41}=0.3\pi$.  
Both powers and currents are fitted with a Pad\'e function $p(\gamma)$ (resp. $\mathcal{J}^{p/E}(\gamma)$)$=(a_0+a_1 \gamma+a_2\gamma^2)/(1-a_3\gamma)$.

%%%%%%%%%
\begin{figure}
\begin{center}
\includegraphics[width=0.9\columnwidth]{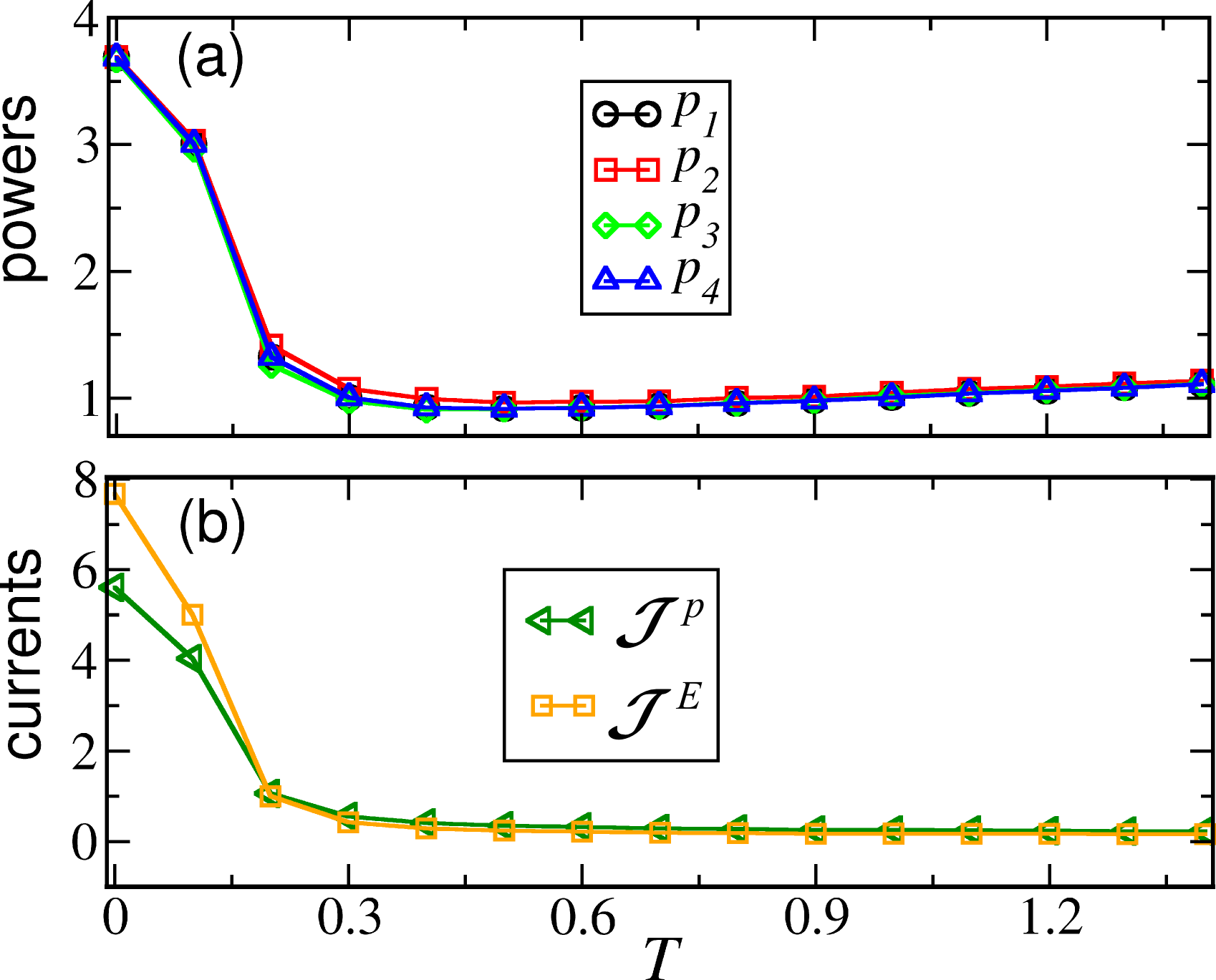}
\end{center}
\caption{(Color online) Oscillator powers (a) and currents (b) vs temperature $T$, computed for $\beta_{41}=0.3\pi$ and $\gamma=0.021$. Powers and currents decrease with temperature $T$, since thermal fluctuations reduce
the coherence between the oscillators. The current $\mathcal{J}^p$ is magnified by a factor 6 for better visibility. The lines are guides to the eye.} 
\label{fig:figure7}
\end{figure}

However, when temperature increases and $\gamma=0.021$ remains fixed, the situation is reversed: powers and currents drop fast and reach an asymptotic value around $T=0.3$. This is most likely due to the fact that temperature does not increase only energy, but also phase fluctuations. Those break the exact synchronisation and tend to suppress transport processes. This feature shows the importance of coherence in energy transport processes of DNLS systems.

%%%%%%%%%%%
\section{conclusions}%
%%%%%%%%%%%
We have demonstrated that the off-equilibrium thermodynamics of a large class of oscillating systems can be described by a $U_1$ lattice gauge theory. The gauge fields correspond to the complex
coupling between the oscillators. The thermodynamical forces that generate circulating currents are the curvature of the fields. Those forces are thus \emph{geometrical} properties of the ensemble (system+reservoirs).

The main prediction of the model, the circulation of persistent currents along close loops, is confirmed by numerical simulations. This result is robust within a large range of temperature of physical parameters and hold far from thermal equilibrium. 
We find that transport increases with chemical potential but is suppressed by thermal fluctuations that tend to destroy the phase coherence of the system, a behaviour coherent with previous studies on the DNLS. 

This work open the path to several possible developments. In particular, the off equilibrium thermodynamics of system with other kind of gauge symmetries \cite{kogut79} could be studied in a similar way. 
The transition between coherent and incoherent transport, where currents are carried mainly by the modulation of the powers and the phase are completely incoherent, should also be investigated.

The role of close paths in our formulation suggests a link with Schnakenberger theory, and further investigation is needed in this direction.

Concerning possible experiments, current circulation could possibly be observed in spin system with the Dzyaloshinskii-Moriya Interaction \cite{zakeri10}, characterised by an asymmetric propagation of spin waves. The dynamics of those systems can be phenomenologically described by means of Eq.(\ref{eq:dnls}) \cite{slavin09} with a complex asymmetric coupling term $A_{mn}\neq A_{nm}$. This could allow for a nonzero anholonomy angle on close paths.

%%%%%%%%%%%
\acknowledgements%
%%%%%%%%%%%
I am grateful to Dr. D. Kuzmanovski and Dr. S. Iubini for useful discussions.

This work was supported by the Swedish Research Council (VR), Energimyndigheten (STEM), the Knut and Alice Wallenberg Foundation, the Carl Tryggers Foundation, the Swedish e-Science Research Centre (SeRC) and the Swedish Foundation for Strategic Research (SSF).

%BIBLIOGRAPHY%
%\bibliography{biblio_gauge}
%\bibliographystyle{plain}{99}
%\bibliographystyle{unsrt}

\end{document}